\documentclass[12pt,a4paper]{article}
\usepackage{graphics,epsfig}

\newcommand{\ba}{\begin{eqnarray}}
\newcommand{\ea}{\end{eqnarray}}
\newcommand{\beqs}{\begin{eqnarray}}
\newcommand{\eeqs}{\end{eqnarray}}

\begin{document}

\textwidth=135mm
 \textheight=200mm
\begin{center}
{\bfseries The structure of nucleons and the description
of the electromagnetic form factors
}
\vskip 5mm
Selyugin O.V.
\vskip 5mm
{\small {\it  Joint Institute for
Nuclear Research, 141980 Dubna, Russia}} \\
\end{center}
\vskip 5mm
\centerline{\bf Abstract}
The comparison of  different sets of PDFs structure functions
    with the description of the whole sets of experimental data
    of  electromagnetic form factors of the proton and neutron is made
    in the frame work of our model [1] of t-dependence of generalized parton distributions (GPDs) 
     and some other models. It is shown that despite a small difference of the description
    of the inelastic processes by the different sets of  PDF there is an essentially
    large difference in the description of electromagnetic form factors of the nucleons.
\vskip 10mm
\section{\label{sec:intro}Introduction}
   The new data of the TOTEM Collaboration 
   show that none of the model predictions can
   describe the elastic cross sections at LHC. 
  One of the main problem of the dynamical models is the form factors of the hadrons.
 In most part, the models are based
  on the assumption that the strong form factors correlate with the electromagnetic form factors. In practice, the models
  use some phenomenological forms of the form factors with the parameters determined by the fit of the experimental data.
  In some works \cite{Miettinen}, the idea was introduce  that the strong form factors can be proportional to the
  matter distribution of the hadrons. In \cite{M1}, the model was developed with the two forms of the form factors
  - one is the exact electromagnetic form factors and the other is proportional to the matter distribution of the hadron.
  Both form factors were obtained from the General Parton distributions (GPDs) which are based on the parton distributions (PDF)
  obtained from the data on the deep inelastic scattering.  The model uses the old PDF obtained in \cite{Martin-02}.
  In the framework of the model, a good description of the high energy  of the proton-proton and proton-antiproton
  elastic scattering was obtained only with 3 high energy fitting parameters. There arrises a question: how the different
  PDF sets describe the electromagnetic form factor of the hadrons. For that, we make a simultaneous fit of  all
   available experimental data on the proton and neutron electromagnetic form factors using the different PDF sets with the
  some model of the $t$-dependence of the GPDs.

\section{The description of the electromagnetic form factors}

The electromagnetic form factors can be represented as the first  moments of GPDs
\ba
 F_{1}^q (t) = \int^{1}_{0}  dx  \ {\cal{ H}}^{q} (x, t); \ \
 F_{2}^q (t) = \int^{1}_{0} dx \  {\cal{E}}^{q} (x,  t),
\ea
following from the sum rules \cite{Ji97}.

 Recently, there were many different proposals for the $t $ dependence of GPDs.
 We introduced a simple form for this
 $t $-dependence~\cite{ST-PRD} based on the original Gaussian form corresponding to that
 of the wave function of the hadron. It satisfies the conditions of non-factorization,
 introduced by Radyushkin, and the Burkhardt condition on the power of $(1-x)^n $
 in the exponential form of the $t $-dependence. With this simple form
  we obtained a good description of the proton electromagnetic Sachs form factors.
  Using the isotopic invariance we obtained good description of the neutron
  Sachs form factors without changing any  parameters \cite{ST-PRD}.

Let us modify the original Gaussian ansatz
 and choose
 the $t$-dependence of  GPDs in the form

\ba
{\cal{H}}^{q} (x,t) \  = && (\frac{2}{3} \ q_u \
     	 exp( 2\alpha_1\frac{(1-x)^{p_1}}{(x_0+x)^{p_2}} t)	  \nonumber \\
     	 && -\frac{1}{3} \ q_d \
        exp( 2\alpha_1 [ \frac{(1-x)^{p_1 k_d}}{(x_0+x)^{p_2}} t)
     	+ d \ x (1-x) ] \ t );
\ea
\ba
{\cal{E}}^{q} (x,t) &&= (\frac{1.673}{uN}\frac{2}{3} \ q_u (1-x)^n) \
     	 exp( 2\alpha_1\frac{(1.-x)^{p_1}}{(x_0+x)^{p_2}} t)	   \\
     	 && -\frac{2.033}{dN}\frac{1}{3} \ q_d \ (1-x)^m)
        exp( 2\alpha_1 [ \frac{(1-x)^{p_1 k_d}}{(x_0+x)^{p_2}} t)
     	+ d \ x (1-x) ] \ t ). \nonumber
\ea
  The parameters $k_d$ and $d$ reflect the possible flower dependence of GPDs.

%
%
   We analyzed four cases: a) with minimum free parameters and flower independence
   (basic variant), as was made in \cite{ST-PRD};
   b) $k$ is taken as a free parameter;
   c) $p_1$ and  $k$ are taken as free parameters;
   d)  $p_1$, $k$ and $d$ are taken as free parameters (maximum flower dependent).
    A further increase in the number of the free parameters does not give  essentially improving of $\chi^2$.
    It leads only to the increasing in the size of the errors of the parameters.
     PDF sets were taken as 15 variants in different works with taking into account the leading order (LO),
      next leading order (NLO) and next-next leading order (NNLO).
      The experimental data on the electromagnetic form factors were represented by 508 points in the maximum
      and 415 points in the minimum variants. The best description was obtained with the PDF sets \cite{Al-09}.
      In this case, all 4 variants of the $t$-dependence gave a very close size of  $\chi^2$.
       Also, we obtained a good description with the PDF sets \cite{Al-12,G-07MSa} and on the third place we can put
       the variants with the PDF sets \cite{Martin-02,M-09-NNLo}.
   It is to be noted that the different PDF sets give similar descriptions of the proton form factors
   and a large difference in the description of the neutron form factors (see Fig.1 and Fig.2).
   Just the neutron data in most part lead to the essentially better description of the polarization data on the
   electromagnetic form factors.

%
%

\begin{figure}
\includegraphics[width=.4\textwidth]{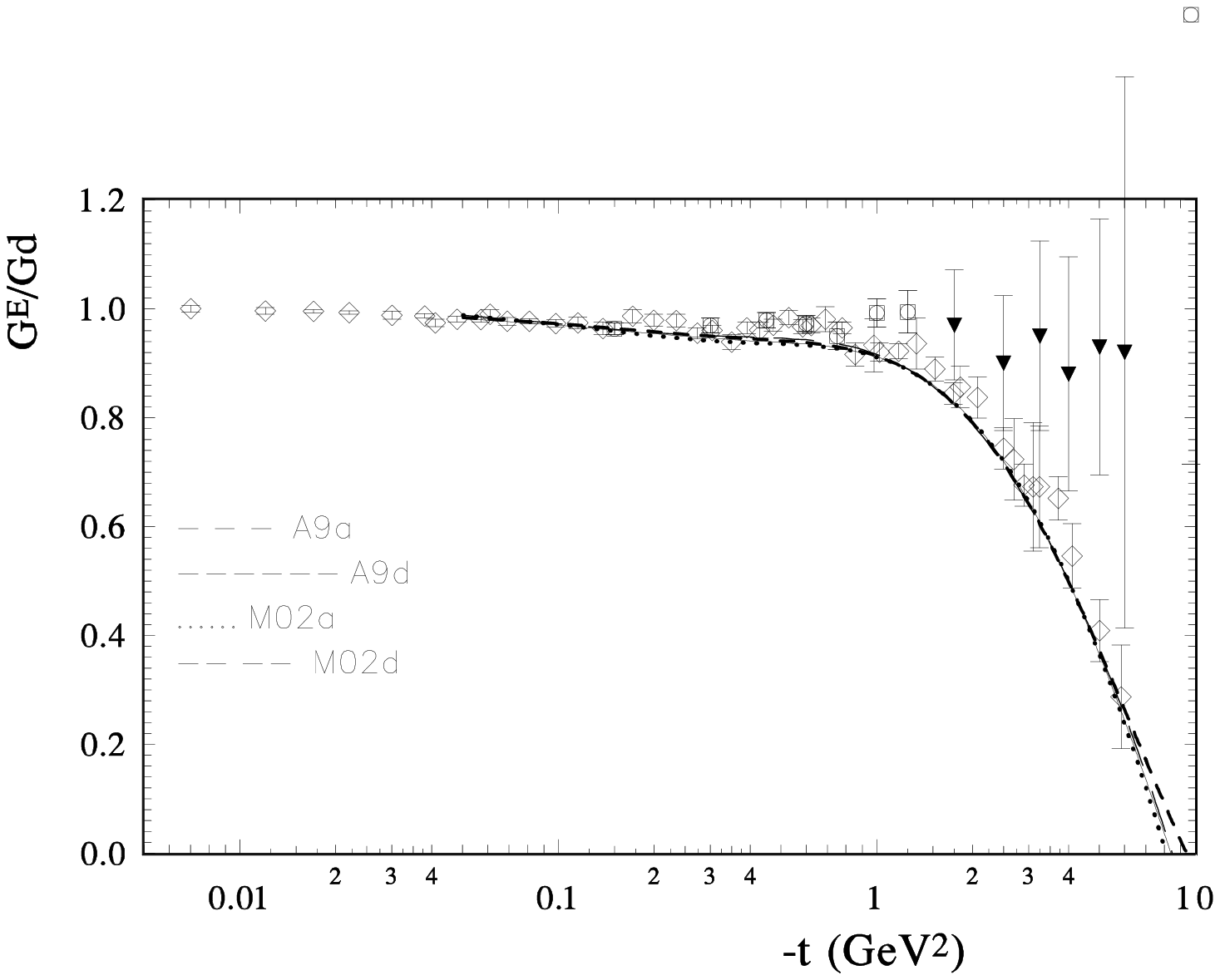} 
\includegraphics[width=.4\textwidth]{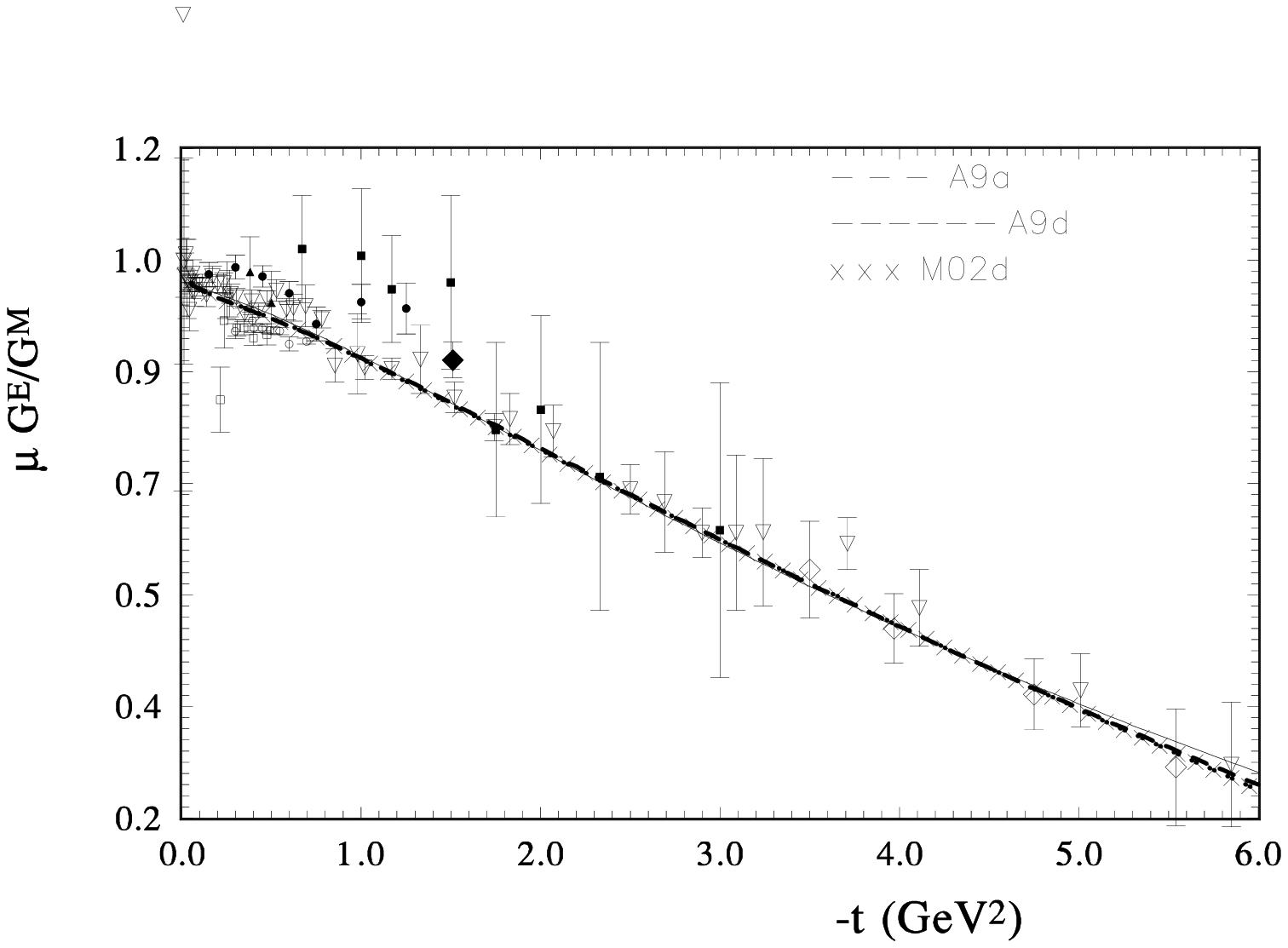} 
\caption{(left)
  The proton $G_{E}/Gd$; (right) $\mu G_e/G_{M}(t)$.
  }
\label{Fig_1}
\end{figure}

\section{Conclusions}
We examined some new forms of the  momentum transfer dependence of the GPDs with
  the different sets of the PDFs. It was found that the new form only slightly differs
  from the simplest $t$-dependence of the GPDs  proposed in \cite{ST-PRD}.
  Contrary to work \cite{Liuty-fd}, we found a small flavor dependence of the $u$ and $d$ components of the GPDs. 
  We found that maximum 5 free parameters are required for a simultaneous
  description of the electromagnetic form factors of the proton and neutron.
  It was shown that the different PDF sets lead to the essentially different descriptions
  of the electromagnetic form factors. The best description was obtained with the
  set of  PDFs \cite{Al-09}.
    In the final analysis we found that a simultaneously description of the proton and neutron
    electromagnetic form factors leads to the "polarization" case of the $t$-dependence of the
    form factors (Fig. 1b). The description of the proton electromagnetic form factors with different PDF sets
     has a large difference only at high $t$.
     The largest difference comes from the description of the electromagnetic form factors of the neutron (see Fig.2).


\begin{figure}
\includegraphics[width=.4\textwidth]{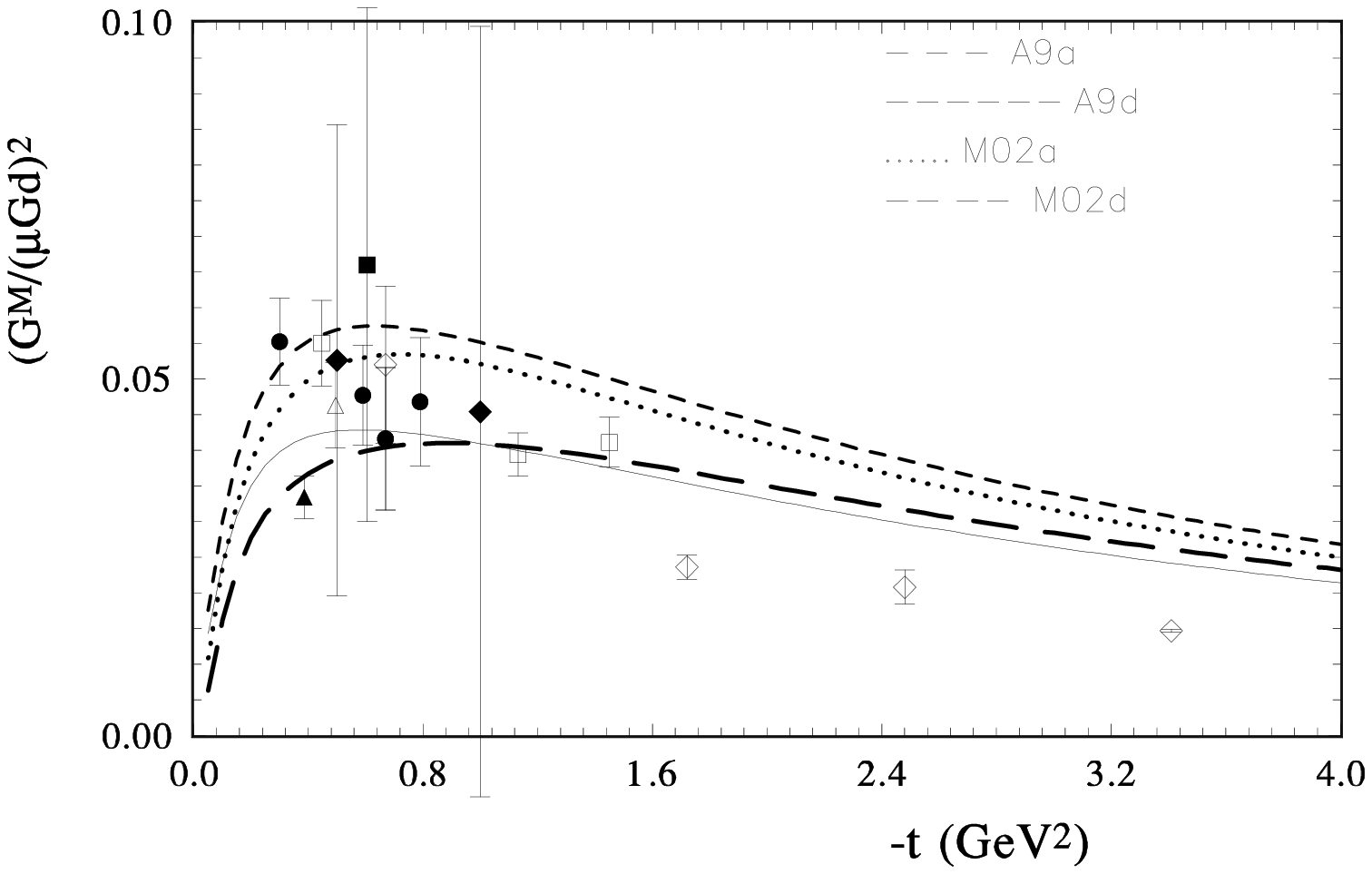} 
\includegraphics[width=.4\textwidth]{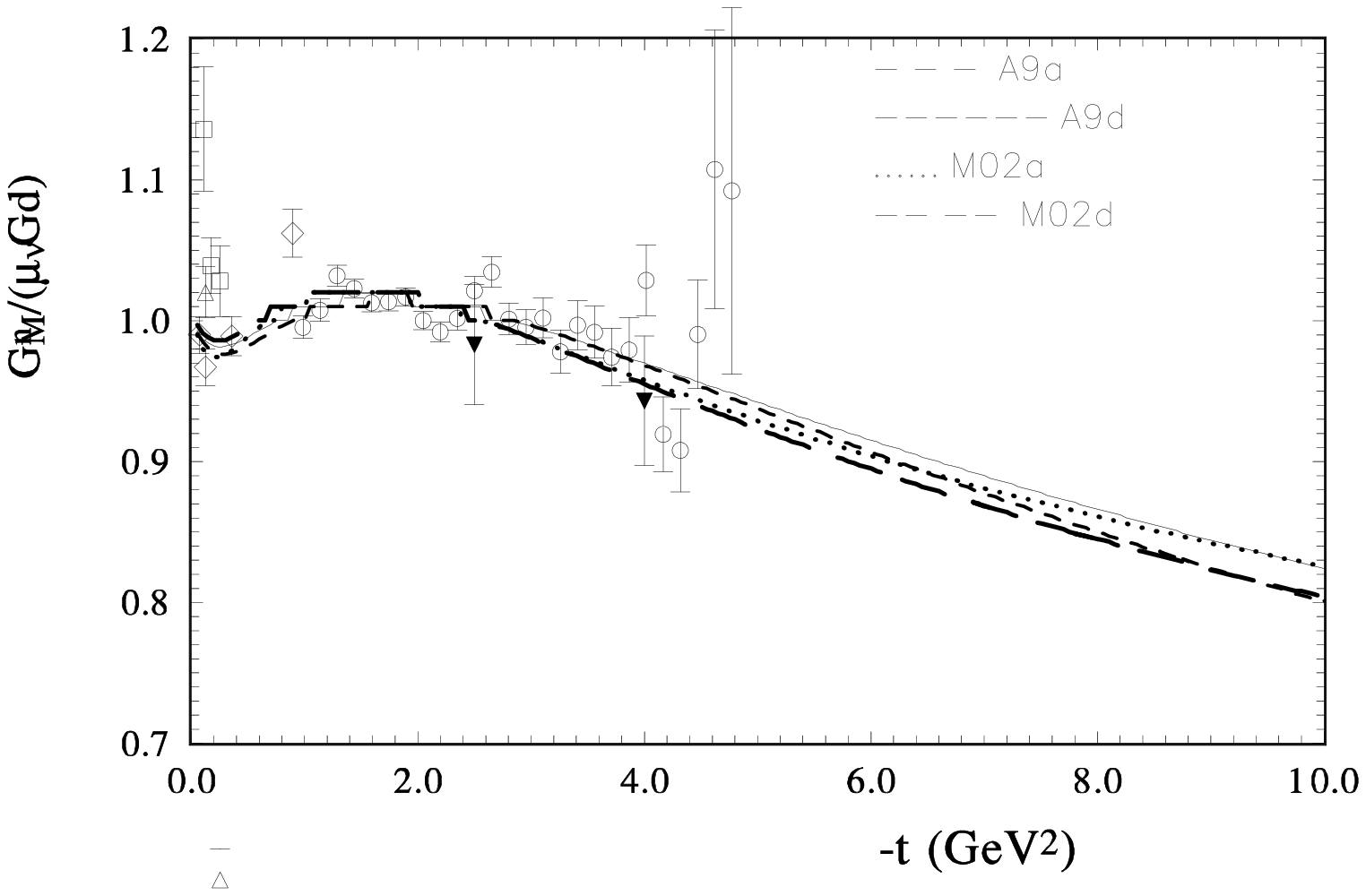} 
\caption{ Neutron $[G_M/(\mu Gd)]^2$ (left - small $t$; (right) - large $t$).
  }
\label{Fig_1}
\end{figure}



{\small
\begin{thebibliography}{99}
{\small


  \bibitem{Miettinen}\textit{ Miettinen  H.} Nucl.Phys. B. 1980. V.166. P.365;
 \textit{Sanielevici S., Valin P.} Phys.Rev. D. 1984. V29. P52.

   \bibitem{M1}
 \textit{Selyugin O.V.} //
 { Eur.\ Phys.\ J. \ C.  2012. V.72. P.2073.}

    \bibitem{Martin-02}   \textit{Martin A.D. et. al.,   } //        {Phys. Lett.} {B. 2002. V.531.} P.216.

  \bibitem{Ji97}
            \textit{Ji X.D.  } //
            {Phys. Lett.} {B. 1997. V.78.} P.610.
  \bibitem{ST-PRD}\textit{Selyugin O.V., Teryaev O.V.} // Phys. Rev. 2008. D.  V.79. P.033003.
%

 \bibitem{Al-09} 
 \textit{Alekhin S. et al.} //  Phys.Rev. D. 2010. V.81. P.014032.
\bibitem{Al-12} 
 \textit{Alekhin S., Bl$\ddot{u}$mlein J., Moch S. } // arXiv:1202.2281.



\bibitem{G-07MSa} 
\textit{M. Gl$\ddot{u}$ck, P. Jimenez-Delgado, E. Reya } // Eur.Phys.J. 2008. C V.53. P.355. 


 \bibitem{M-09-NNLo} \textit{
     Martin A.D. et al. } //        {Phys. Lett.} {B. 2009. V..} P..


\bibitem{Liuty-fd}\textit{J. Osvaldo Gonzalez-Hernandez, Simonetta Liuti, Gary R. Goldstein, Kunal Kathuria} //arXiv:1206.1876.


}

\end{thebibliography}
}
\end{document}